\documentclass[fleqn,usenatbib]{mnras}

\usepackage{newtxtext,newtxmath}

\usepackage[T1]{fontenc}

\DeclareRobustCommand{\VAN}[3]{#2}
\let\VANthebibliography\thebibliography
\def\thebibliography{\DeclareRobustCommand{\VAN}[3]{##3}\VANthebibliography}

\usepackage{graphicx}	
\usepackage{amsmath}	
\usepackage{CJK}

\defcitealias{kimmig2020}{K20}

\title[Planet Migration in Windy Discs]
{Planet Migration in Windy Discs}

\author[Wu \& Chen 2024]
{Yinhao Wu (吴寅昊),$^{1}$
Yi-Xian Chen (陈逸贤)$^{2}$
\\
$^{1}$School of Physics and Astronomy, University of Leicester, Leicester LE1 7RH, UK \href{mailto:email@domain}{yw505@leicester.ac.uk} (YW)\\
$^{2}$Department of Astrophysical Sciences, Princeton University, USA \href{mailto:email@domain} {yc9993@princeton.edu} (YC)\\
}

\date{Accepted XXX. Received YYY; in original form ZZZ}

\pubyear{2024}

\begin{document}
\begin{CJK*}{UTF8}{gbsn} 
\label{firstpage}
\pagerange{\pageref{firstpage}--\pageref{lastpage}}
\maketitle

\begin{abstract}
Accretion of protoplanetary discs (PPDs) 
could be driven by MHD disc winds rather than turbulent viscosity. 
With a dynamical prescription for angular momentum transport induced by disc winds, 
we perform 2D simulations of PPDs to systematically investigate the rate and direction of planet migration in a windy disc. 
We find that the the strength of disc winds influences the corotation region  
similarly to 
the "desaturation" effect of viscosity. 
The magnitude and direction of torque 
depend sensitively on the hierarchy between the radial advection timescale across the horseshoe due to disc wind $\tau_{\rm dw}$, 
the horsehoe libration timescale $\tau_{\rm lib}$ and U-turn timescale $\tau_{\rm U-turn}$.   
Initially, as  
wind strength increases and the advection timescale shortens, 
a non-linear horseshoe drag emerges when $\tau_{\rm dw} \lesssim \tau_{\rm lib}$, which tends to drive strong outward migration. 
Subsequently, 
the drag becomes linear and planets typically still migrate inward when $\tau_{\rm dw} \lesssim \tau_{\rm U-turn} \sim \tau_{\rm lib}h$, 
where $h$ is the disc aspect ratio. 
For a planet with mass ratio of $\sim 10^{-5}$, the zone of outward migration sandwiched between inner and outer inward migration zones corresponds to $\sim $ 10-100 au in a PPD with accretion rates between $10^{-8}$ and $10^{-7}$ $M_\odot \text{yr}^{-1}$.
\end{abstract}

\begin{keywords}
planet-disc interactions -- protoplanetary discs -- planets and satellites: formation 
\end{keywords}



\section{Introduction}

Orbital migration plays an important role in the formation of planets, especially for super-Earths and gas giants, ultimately shaping the architecture of planetary systems. 
Migration of a planet arises from its tidal interaction with the protoplanetary disc (PPD) \citep{Goldreich1980,LinPapaloizou1986,KleyNelson2012,Pablo-2024review}, 
which could cause a gap to open up in the disc when the planet mass is large enough \citep{1986ApJ...307..395L,Kanagawa2015}.  Although most work assumes, for convenience, that the planets are fixed in their orbit, 
the observational signatures created by migrating planets can differ from those associated with stationary planets \citep{Wu2023migration,Wu-SKA-2024}. 
Extensive studies have examined this process within the context of discs with accretion driven by turbulence,
often modeled as a viscosity term in numerical simulations. 
These studies conclude simplified but useful formulae for planet migration torques \citep{Paardekooper2010, Kanagawa2018}, applicable to locally isothermal and laminar viscous discs,
which are implemented in population synthesis of planetary systems \citep{ida2008,Ida2018}.  Nevertheless, 
there remain uncertainties in the migration process due to several factors. 
These factors include thermal torques in non-isothermal discs \citep{Guilera2021, Yun2022}, 
dust feedback effects \citep{Llambay2018, Guilera2023}, 
realistic turbulent viscosity \citep{Nelson2005,BaruteauLin2010,Wu2024chaotic}, 
and sharp density gradients at gap edges \citep{Chen2020},
most of which are not fully captured by existing prescriptions. 

On the other hand, 
measurements of velocity dispersions from CO lines 
in most PPDs 
suggest the absence of strong turbulence \citep{Flaherty2017,Flaherty2020}.
This observation is consistent with 
non-ideal Magneto-hydrodynamic (MHD) simulations of PPDs 
where strong MRI turbulence is 
often suppressed due to poor ionization \citep{2013BaiStone}. Instead, 
a magnetocentrifugal wind 
is launched from the upper surfaces of PPDs,  
which drives an inward accretion \citep{Simon2013,Armitage2013,Suzuki2016,Cui2021}. 

Detailed outcome of planet-disc interaction 
in a "windy disc", i.e., a PPD with accretion driven by MHD disc winds, has been investigated with non-ideal MHD simulations by \citet{AoyamaBai2023,Lesur2023,Hu2024}. 
In most simulations, the torque acting on the planet by the surrounding disc gas drives an inward migration, 
although the planet is able to concentrate poloidal magnetic flux in its vicinity, 
causing velocity and density distribution in the corotation region to be highly asymmetric. 
However, these studies deal with gap-opening planets, 
since it becomes computationally expensive to extrapolate down to sub-thermal companions \footnote{ typically planets with masses below $40 \ M_\oplus$ for disc aspect ratio $h\sim 0.05$.}  while capturing dynamics within their Bondi radii with the same relative resolution.

Taking a more general perspective, 
even without resolving the details of magnetic fields, 
key aspects of the migration process 
can still be captured 
by examining the phenomenological 
dynamical effects of wind.
For instance, 
\citet[][hereafter \citetalias{kimmig2020}]{kimmig2020} conducted a series of 2D hydrodynamical simulations of migrating giant planets. 
They simulate wind accretion by imposing a monotonic inward radial flow of gas \citep{BlandfordPayne1982, Ogihara-2015, Ogihara-2018}, 
and discover that planets tend to migrate outward when wind is strong. 
Specifically, this transition from inward to outward migration occurs when the radial advection timescale across the horseshoe region, 
$\tau_{\rm dw}$, becomes shorter than the libration timescale $\tau_{\rm lib}$. \citetalias{kimmig2020} related this transition to the 
asymmetry of horseshoe caused by the amount of gas flowing into the inner disc before librating to the lower half of the horseshoe. 
In other words, an excess of materials in the upper horseshoe region is caused by quick advection of material across the horseshoe region.
Although not explicitly mentioned, 
this is similar to how moderate viscosity could generate a non-linear outward corotation torque/horseshoe drag in classical Type I migration \citep{Masset2001, PaardekooperPap2009,Baruteau2013}, 
which occurs at $\tau_{\rm vis} \lesssim \tau_{\rm lib}$, 
where $\tau_{\rm vis} $ 
is the viscous \textit{diffusion} timescale across the horseshoe region.
Due to the different dependencies of $\tau_{\rm dw}$ and $\tau_{\rm vis}$ on disc aspect ratio $h$,
achieving the same transition 
in a windy disc requires a significantly higher accretion rate. 
Nevertheless, 
it is important to note that $\tau_{\rm vis}$ here simply represents a spatially and temporally averaged \textit{net} radial transport timescale, 
assuming a reasonable surface density gradient has been established across the disc. 
While this diffusion process yields a similar effect to radial inflow, the underlying physics differ. 

However, for classical migration 
in viscous discs, 
this enhancement of horseshoe drag with respect to viscosity is not monotonic. 
Instead, 
the outward horseshoe drag will relax to the linear corotation torque, which is smaller than its maximum non-linear value by order unity, 
in the limit where $\tau_{\rm vis}$ becomes even shorter than the U-turn timescale $ \tau_{\rm U-turn} \sim \tau_{\rm lib} h$ \citep{BaruteauMasset2008,Baruteau2013}. 
A physical interpretation of this transition 
is that material 
might not even have time to linger in one side of the horseshoe, before diffusing into the inner disc. 
This puts a question to whether the outward migration observed in \citetalias{kimmig2020} 
maintains at even larger wind accretion rates or arbitrarily short $\tau_{\rm dw}$.

In this work, 
we perform 2D hydrodynamic simulations with a dynamical prescription for wind-driven accretion and extend the 
parameter space of \citetalias{kimmig2020} to low mass planets. 
We find that the horseshoe drag 
will indeed relax to a linear value and 
planets are typically in a state of inward migration if $\tau_{\rm dw}\lesssim \tau_{\rm U-turn}$. 
In other words, 
there is an upper limit for the wind accretion parameter that allows for outward migration. 
The letter is organized as follows: \S 2 outlines our numerical approach for simulating planet migration in a windy disc; \S 3 presents our results and intepretations and \S 4 discusses the implications of our parameter survey.

\begin{figure}
\centering
\includegraphics[width=0.99\hsize]{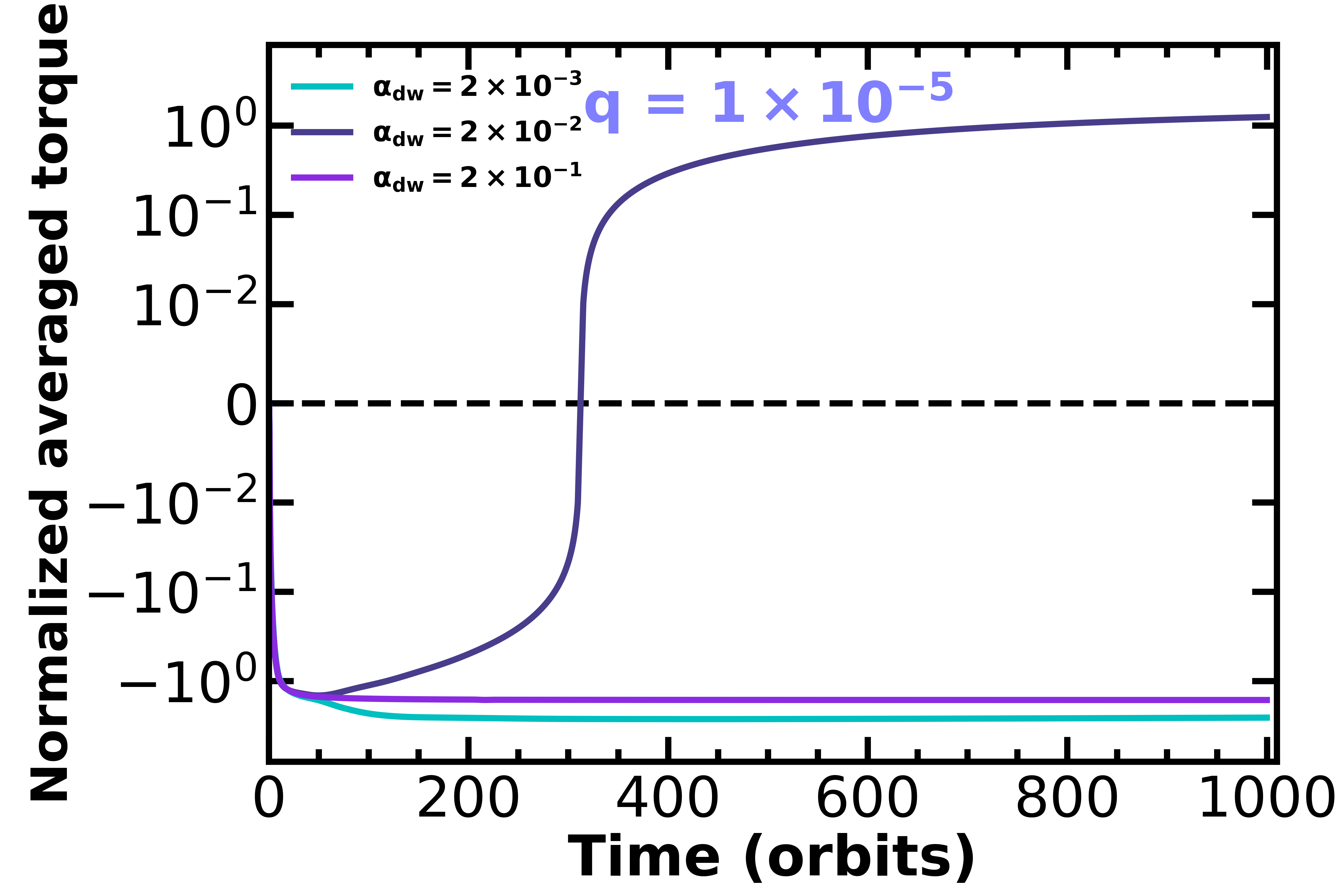}
\caption{The running-time average migration torque (normalized by $\Sigma_0 (q/h_0)^{2}r_0^4 \Omega_0^2$) for our fiducial cases. 
The planet to stellar mass ratio is 
$q=1\times10^{-5}$, i.e., $3.3 M_{\oplus}$ for $1 M_{\odot}$, 
and three different wind accretion parameters are chosen. 
The black dashed line indicates where the averaged torque equals zero. The orbital periods represented on the time axis are based on the planet's initial position, i.e., $r_0$.}
\label{fig:rtatorque_3cases}
\end{figure}

\begin{figure}
\centering
\includegraphics[width=0.8\hsize]{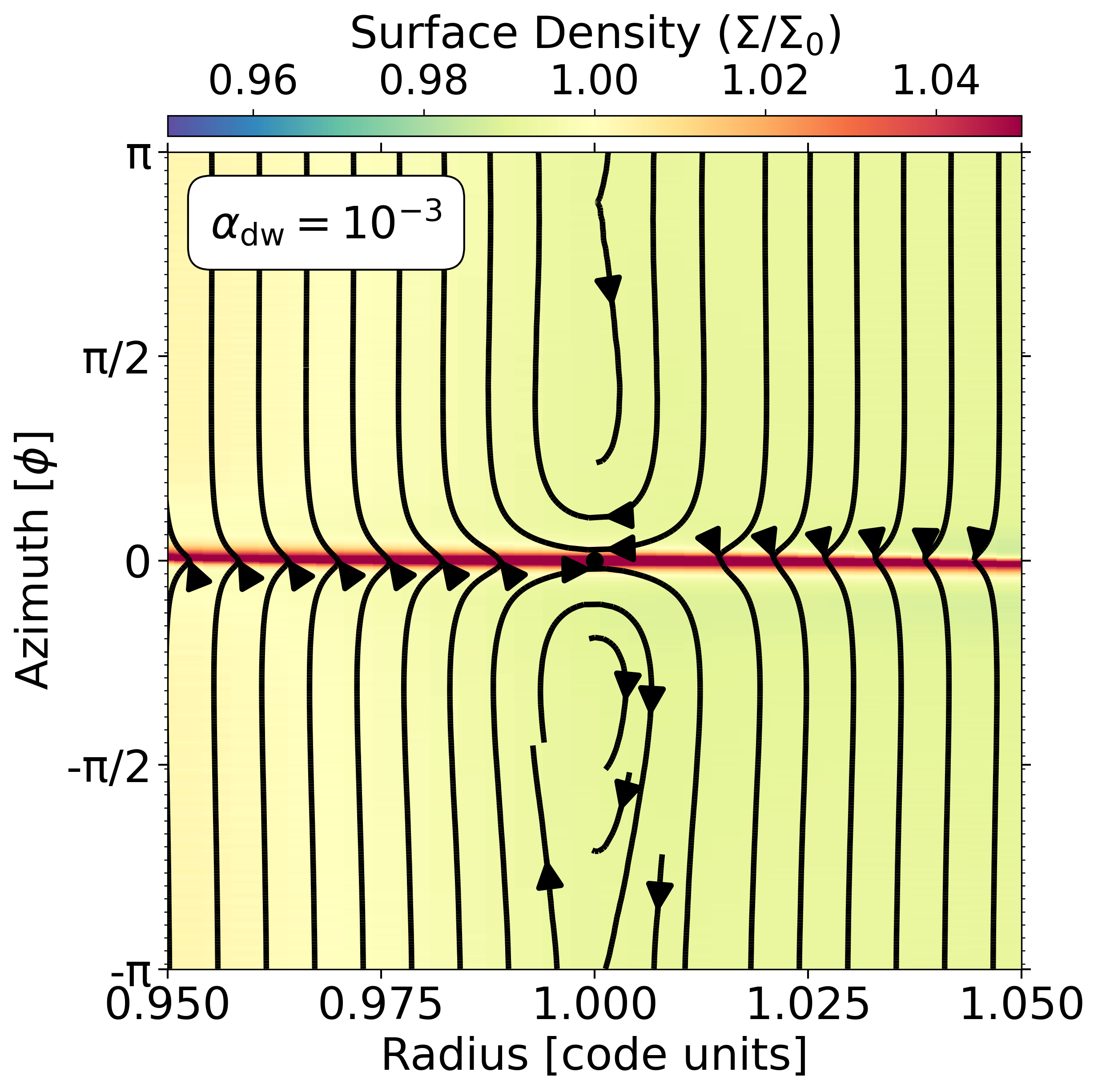}
\includegraphics[width=0.8\hsize]{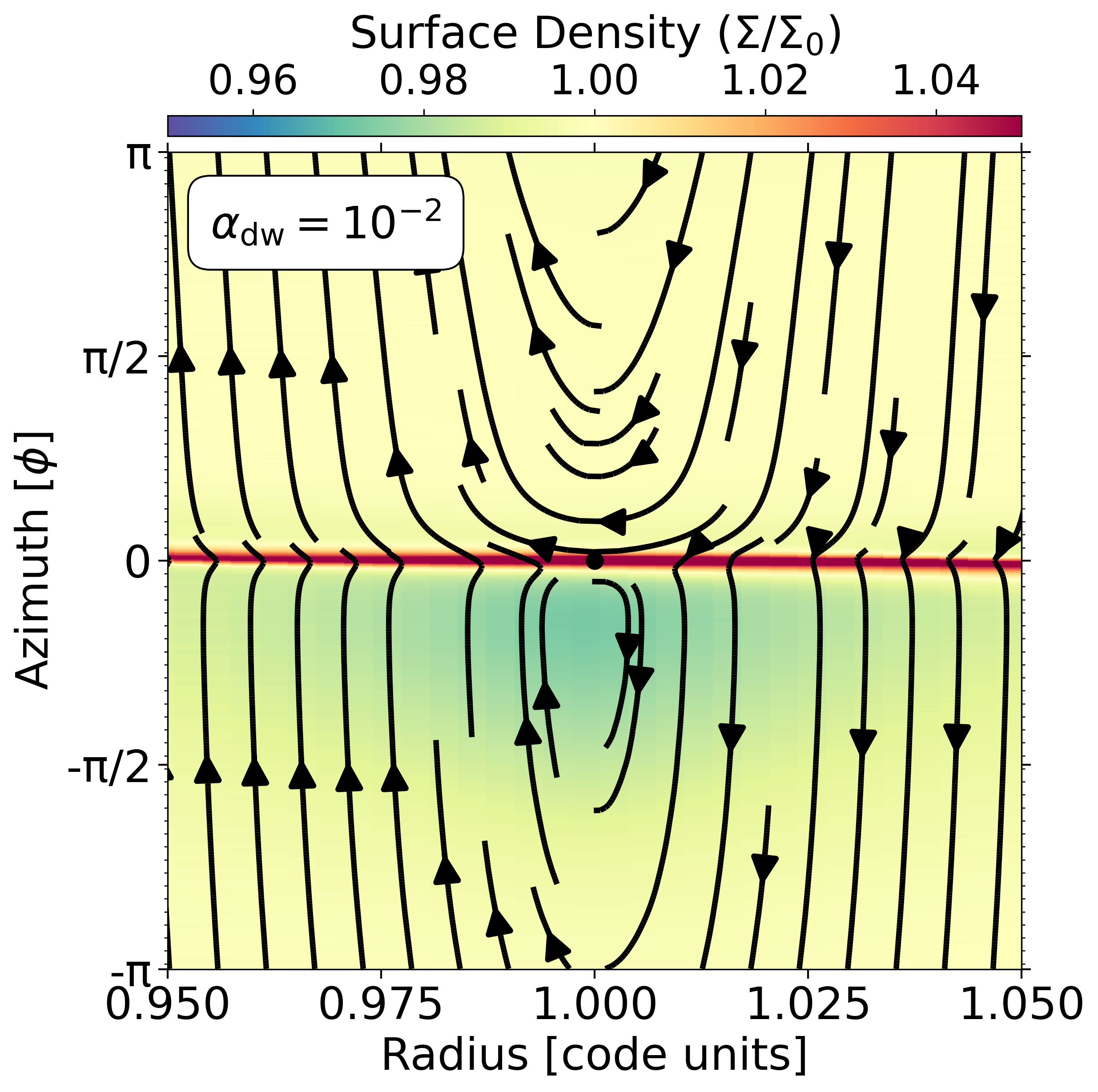}
\includegraphics[width=0.8\hsize]{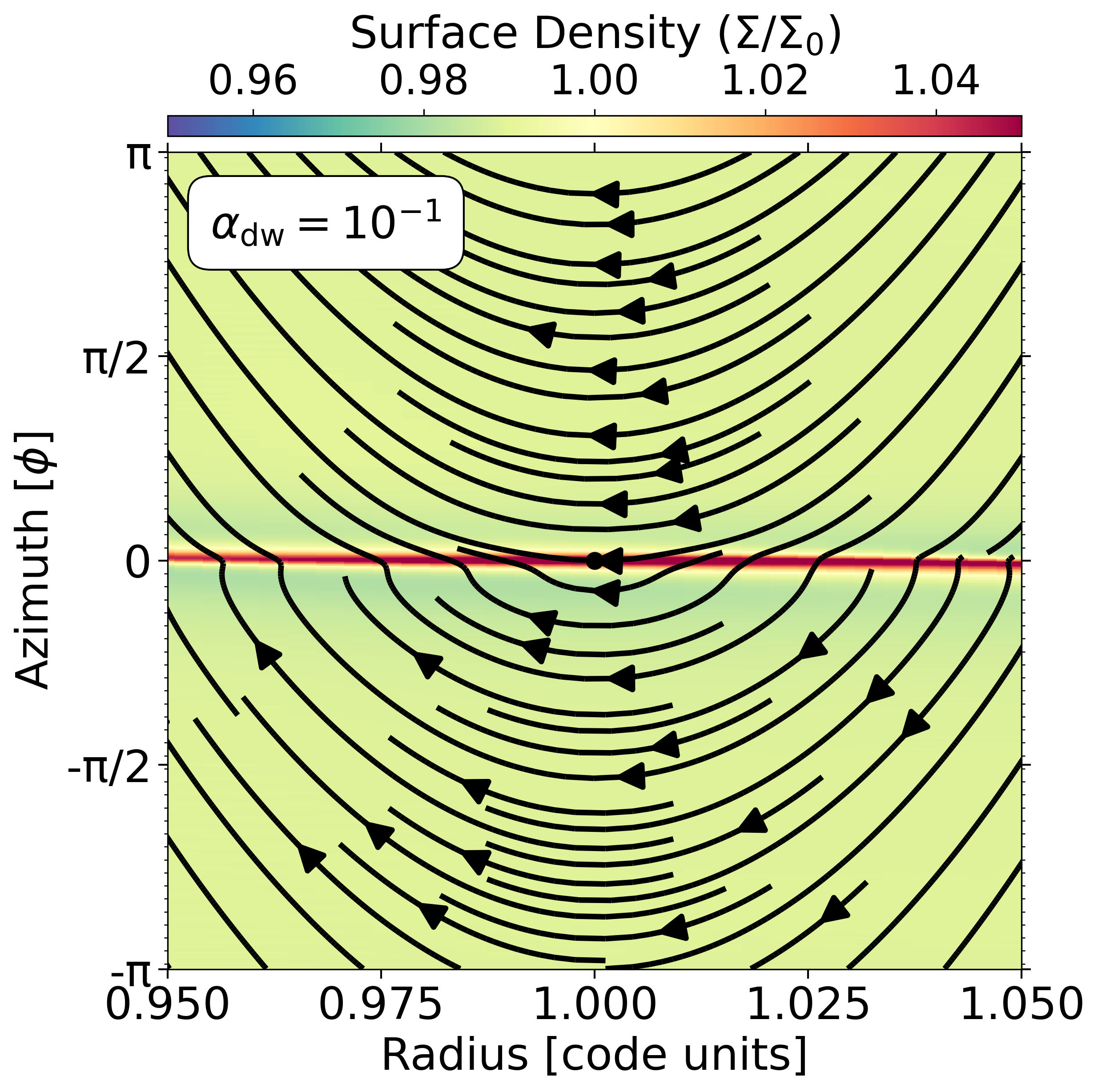}
\caption{Normalized 
surface density maps (colorbar shown above) and streamlines 
for the reference cases from Figure \ref{fig:rtatorque_3cases} 
(wind accretion parameter $\alpha_{\rm dw}$ is indicated in each panel). The position of the planet is marked with a black dot. Note that only the middle panel shows significant azimuthal imbalance of density across the horseshoe region, which leads to a strong negative migration torque. }\label{fig:streaming_line_alpha_dw}
\end{figure}


\section{Numerical Setup}
To study the effect of disc wind (or in general, an imposed radial advection velocity toward the star) 
on planet migration, 
we use a modified version \citep{WU_Chen_jiang_2023, Wu2024DMTau} of the grid-based hydrodynamic code \texttt{FARGO3D} \citep{FARGO3D} 
with a dynamical disc wind prescription given by \citet{Tabone22}, 
and extended to 2D in \citet{MHD-wind-Elbakyan}: 
\begin{equation}
    \Gamma = \frac{1}{2} \sqrt{ \frac{GM_*}{r}} V_{\rm dw}\;,
    \label{Gamma}
\end{equation}
where $M_*$ is the mass of the host star , $G$ is the gravitational constant, and $V_{\rm dw}$ is the characteristic radial velocity of gas induced by MHD disc winds \citep{Tabone22}:
\begin{equation}
        V_{\rm dw} = -\frac{3}{2}\alpha_{\rm dw} h^2 V_{\rm K}\;,
    \label{v_dw}
\end{equation}
here $\alpha_{\rm dw}$ is a dimensionless parameter to describe the strength of the disc wind, 
defined in a similar manner to the turbulent viscosity parameter $\alpha_{\rm v}$ \citep{SS1973}, $h=H/r$ is the disc aspect ratio, 
and $V_{\rm K}$ is the Keplerian velocity at $r$.
Our $\alpha_{\rm dw}$ can be simply related to the mass-loss parameter $b$ of \citetalias{kimmig2020} by $b\sim \pi \alpha_{\rm dw} h^2$. \footnote{Assuming that the magnetic lever arm parameter $(r_{\rm A}/r_{\rm F})^2 -1$ is of order unity, where $r_{\rm A}$ and $r_{\rm F}$ are the Alfv\'{e}n point and the footpoint of the magnetic field line.}
Note that the truth of angular momentum transport in PPDs may involve the interaction between MHD disc winds and turbulent viscosity \citep{WU_Chen_jiang_2023,Wu2024DMTau}. However, in order to systematically study the effects of MHD disc winds, we set the Shakura-Sunyaev  viscosity parameter to $\alpha_{\rm v} = 0$ in our simulations, leaving behind only a negligible amount of numerical viscosity.

Similar to viscous discs, the relationship between the accretion rate $\dot{M}_{\text{acc}}$ and $\alpha_{\rm dw}$ can be expressed as \citep{Tabone22}:
\begin{equation}
        \dot{M}_{\rm acc} = \frac{3 \pi \Sigma \alpha_{\rm dw} c_s^2}{\Omega},
    \label{Mdot}
\end{equation}
where $c_s$ is the sound speed, $\Omega$ is the Keplerian orbital frequency around the central star.

The relationship between wind-driven accretion rate $\dot{M}_{\rm acc}$ and the magnetic field strength $B_{\rm mid}$ at the midplane of PPDs can be parameterised by \citep{Bai-Goodman-2009, Weiss-Bai-Fu-2021}:


\begin{equation}
    B_{\rm mid} \simeq m \left(\dfrac{2}{\sqrt3}\dfrac{\dot{M}_{\rm acc} \Omega }{r} \right)^{1/2} f'^{1/2},
    \label{B_Mdot}
\end{equation}
here $m$ and $f'$ are two free parameters related to the configuration of magnetic fields, 
while the scaling over $\dot{M}$ and $r$ is obtained by requiring magnetic torque to drive some fraction of the total 
angular momentum loss \citep[][see their Equation 7]{Bai-Goodman-2009}. This translates to

\begin{equation}
     B_{\rm mid} \simeq m \left(\dfrac{2}{\sqrt3}\dfrac{3\pi \Sigma \alpha_{\rm dw} c_s^2}{r} \right)^{1/2} f'^{1/2}.
     \label{B_alpha}
\end{equation}
We will utilize these scalings to relate our parameter space to physical conditions in \S \ref{sec:Psurvey}.


For the disc model, we set the initial gas surface density profile to be
\begin{equation}
    \Sigma_{\rm g}(r) = \Sigma_0 ({r}/{r_0})^{-1}\;,
    \label{sigma-disc0}
\end{equation}
here $\Sigma_0 = 2 \times 10^{-4} M_*/r_0^2$ at the planet's initial orbital radius $r_0$. And the disc aspect ratio is
\begin{equation}
h(r) = c_s/V_{\rm K} = H/r = h_0 (r/r_0)\;,
\end{equation}
where $h_0=0.05$ at $r=r_0$, which implicitly sets the vertically isothermal structure of the disc.

We apply the wave-damping boundary conditions, 
activating the
\texttt{KEPLERIAN2DDENS} and \texttt{STOCKHOLM} option of \texttt{FARGO3D}.
This treatment is similar to \citetalias{kimmig2020} and our previous 2D simulations in \citet{MHD-wind-Elbakyan,WU_Chen_jiang_2023}. 
We also take into account the indirect terms arising from the planet and the disc accelerations (i.e., the option GASINDIRECTTERM of \texttt{FARGO3D}), 
since the star remains at the frame's origin.

In our simulations, the computational domain ranges from $0.3\,r_0$ to $2.0\,r_0$ radially, where $r_0$ is the code unit length and the initial planet orbital radius. 
The simulation domain is resolved by 512 grid cells in the radial direction spaced logarithmically and 1536 grid cells in the azimuthal direction. 
We run all our simulations to 1000 orbits at the planet's initial orbital radius $r_0$, by which time the running-time average torques have typically reached a steady state. 
In most of the simulations, 
we keep the planet on a fixed orbit rather than allowing it to migrate freely, 
as this allows for better scale-free analysis of the torque. 
We will also present a few cases where this assumption is relaxed to demonstrate the applicability of our results to freely migrating planets.

\section{Results}

\subsection{Fiducial Case}

Figure \ref{fig:rtatorque_3cases} shows the running-time average of total migration torque, 
normalized by $\Sigma_0 (q/h_0)^2  \Omega_0^2 r_0^4$, 
in our fiducial simulations. 
The planet mass ratio is $q = 10^{-5}$, considerably below the thermal mass ratio $h_0^3$ which is not enough to open a gap in the disc, and three representative wind accretion parameters are chosen. 

Comparing the $\alpha_{\rm dw} = 2\times 10^{-3}$ and $\alpha_{\rm dw} = 2\times 10^{-2}$ cases, 
there is indeed a reversal of the direction of migration after reaching a steady-state at higher wind accretion rate. 
This is expected to occur when $\alpha_{\rm dw}$ exceeds the value of $10^{-2}$, or $\tau_{\rm dw} < \tau_{\rm lib} $, where 
\begin{equation}
    \tau_{\rm dw} = x_s/V_{\rm dw}
\end{equation}
is the time-scale to cross the horseshoe region, 
with typical width $x_s\sim \sqrt{q/h} r_0$, 
while the libration timescale is given by
\begin{equation}
\tau_{\rm lib} = \frac{4 \pi r_p}{(3 / 2) \Omega_p x_s}.
\end{equation}
The direct interpretation of the transition from inward to outward migration is the following: 
when $\tau_{\rm dw} < \tau_{\rm lib} $, 
a considerable fraction of gas captured onto the horseshoe orbit
can now advect into the inner disc before completing one horseshoe libration, 
leading to an excess of gas density in the upper horseshoe compared to the lower horseshoe \citepalias{kimmig2020}. 
This can be seen from the density distribution in the first two panels of Figure \ref{fig:streaming_line_alpha_dw}. 


This first transition, as studied in \citetalias{kimmig2020}, is similar to the classical argument of corotation torque desaturation in a viscous disc \citep{PaardekooperPap2009,Baruteau2013}. 
The major difference is that 
$\tau_{\rm lib}$ in that case is compared to the viscous diffusion timescale $\tau_{\rm vis} = x_s^2/\nu, \nu = \alpha_{\rm v} h^2 \Omega r^2$. 
When $\tau_{\rm vis}\lesssim \tau_{\rm lib}$, 
a considerable fraction of gas can on average diffuse 
into the inner disc before it has time to complete one libration and smooth the vortensity profile, and outward horseshoe drag will reach its maximum non-linear value, significantly reducing inward migration rate or even shifting the direction of migration. 
However, in the limit of strong viscous diffusion, 
the total migration torque relaxes to the value determined by the summation of linear Lindblad and corotation torques, 
which ultimately drives inward migration. 

Similarly, in windy discs, we should not expect the direction of migration to remain outward at arbitrarily high $\alpha_{\rm dw}$. 
As we can see from Figure \ref{fig:rtatorque_3cases}, for our simulation with $\alpha_{\rm dw} = 0.2$, 
the average total torque in steady-state is again negative. 
From the density distribution 
shown in the bottom panel of Figure \ref{fig:streaming_line_alpha_dw}, 
the azimuthal asymmetry of the horseshoe region becomes insignificant. 
This is because the wind advection timescale is 
even shorter than the U-turn timescale,
therefore material cannot even 
stay in the upper half of the horseshoe region long enough to produce an effective 
mass imbalance.  The radial components of the streamlines are so large that nearly everything is circulating, with the remaining librating region compressed to a very small azimuth range, or a tiny "libration island" below the planet.

\begin{figure}
\centering
\includegraphics[width=0.99\hsize]{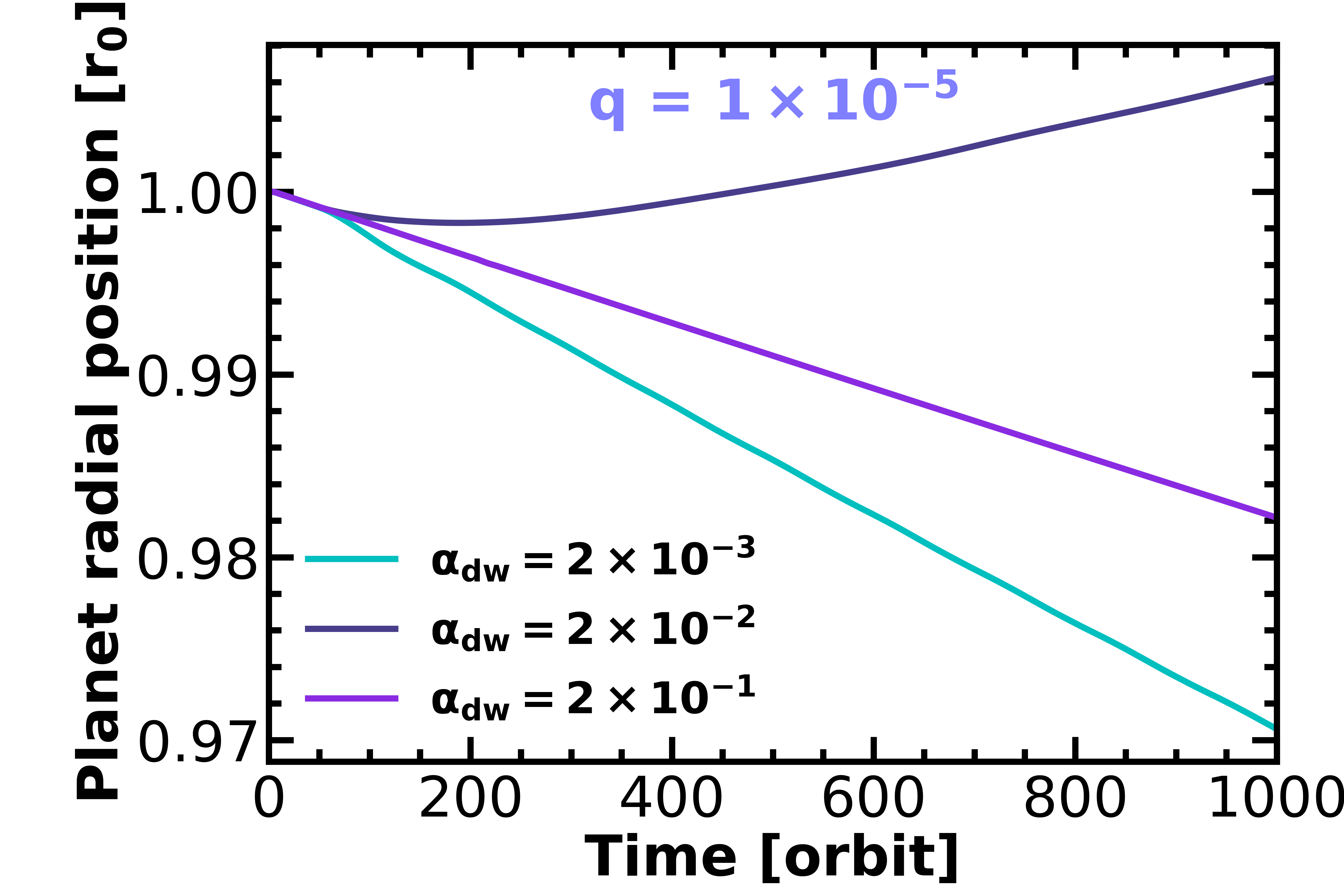}
\caption{The time evolution of 
the planet's radial position in our additional simulations. 
The parameters are the same as our fiducial cases shown in Figure \ref{fig:rtatorque_3cases}，
but we let the planet orbit 
freely evolve under the effect of disc torque.}\label{fig:migration}
\end{figure}

For these fiducial parameters, we also perform simulations where we let the planet's orbit evolve. Figure \ref{fig:migration} shows the semi-major axis evolution over time for these simulations up to 1000 orbits at the planet's original orbital radius.
In this scenario, the planet's overall orbital evolution is expected to be affected by additional factors such as dynamical torques and runaway migration \citep{MassetPapaloizou2003,Paardekooper2014}. 
Although the interaction 
between these effects and the intrinsic torque from the unbalanced horseshoe drag may be complex, 
the steady-state migration rates 
at the end of the simulations are consistent with those observed in the corresponding fix-planet runs.

\subsection{Parameter survey}
\label{sec:Psurvey}

\begin{figure}
\centering
\includegraphics[width=0.99\hsize]{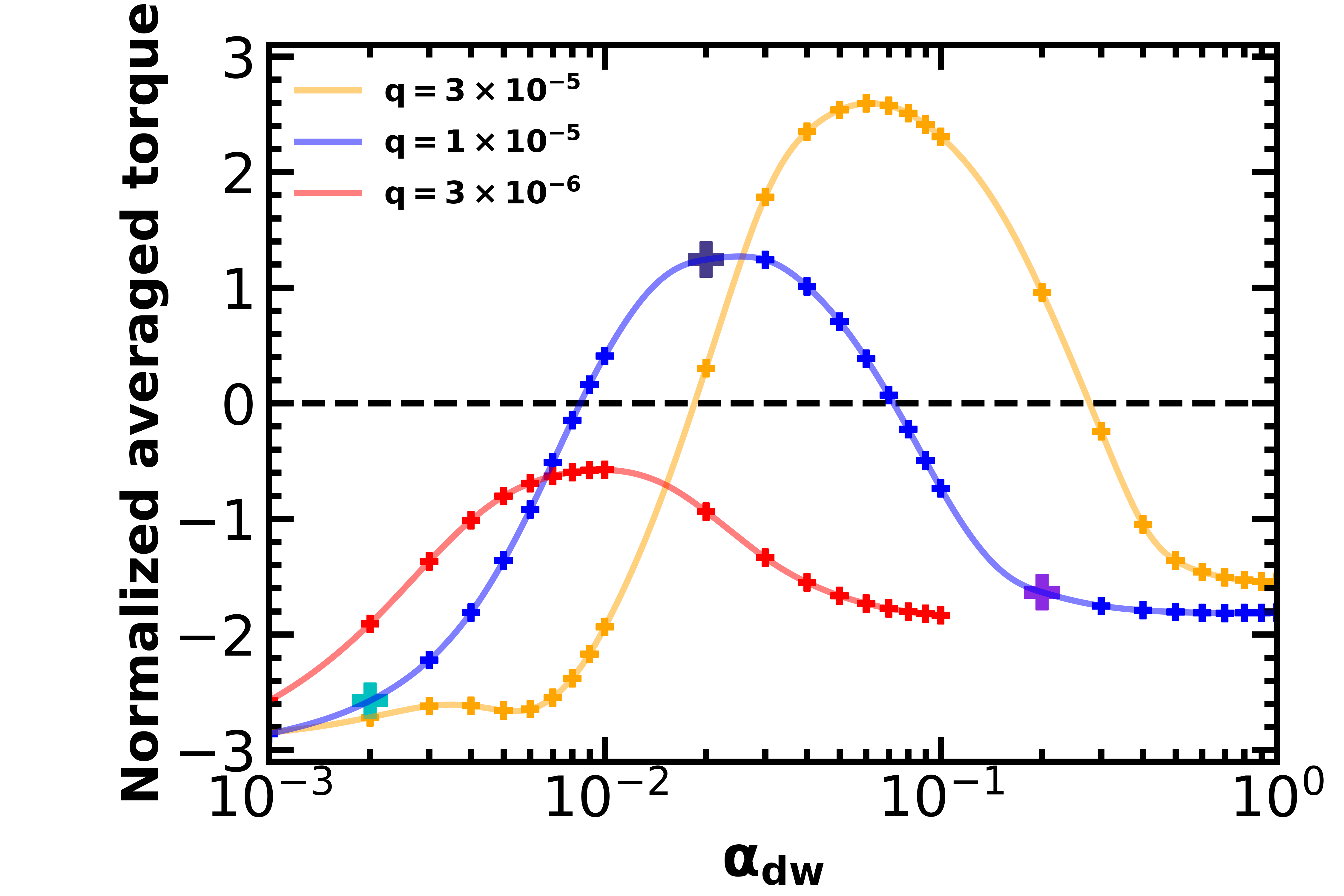}
\caption{Normalized migration torque (normalized by $\Sigma_0 (q/h_0)^{2}r_0^4 \Omega_0^2$) measured in steady states as a function of wind accretion parameter $\alpha_{\rm dw}$. 
If the value of the torque is positive, it indicates that the planet is migrating outward; if the value is negative, the planet is migrating inward.
The solid lines of different colors represent different planet mass ratio $q$, it was fitted by using the \texttt{Python} smoothing interpolation function. 
The black dashed line indicates where the averaged torque equals zero. 
Plus signs indicate data points and solid lines are smoothed profiles.
The cases marked with three enlarged plus signs in cyan ($\alpha_{\rm dw}=2\times10^{-3}$), dark slate blue ($\alpha_{\rm dw}=2\times10^{-2}$), 
and blue violet ($\alpha_{\rm dw}=2\times10^{-1}$) correspond to the fiducial cases shown in Figure \ref{fig:rtatorque_3cases} and  \ref{fig:streaming_line_alpha_dw}.}
\label{fig:rtatorque_radial}
\end{figure}

The steady-state torque measurements from the rest of our simulations are summarized in Figure \ref{fig:rtatorque_radial}.
For three values of planet mass $q$, 
we perform 20 
(for $q=3\times10^{-6}$) 
or 30 (for other) simulations each to cover transitions over the $\alpha_{\rm dw}$ parameter space. 
If we define $M_*=1M_{\odot}$ and $r_0 = 5$ au for normalisation, then $\Sigma_0 \approx 70g/cm^{2}$ at 5 au, which is the same as \citetalias{kimmig2020}.
The strength of the disc winds in our parameter space 
$\alpha_{\rm dw}\in[10^{-3},1]$ would correspond to a range of accretion rates $\dot{M}_{\rm acc}\in[2\times10^{-9}, 2\times10^{-6}]$ $M_\odot\rm yr^{-1}$ given by Equation \ref{Mdot}.

Based on recent simulations incorporating all non-ideal MHD effects \citep[e.g.,][]{AoyamaBai2023}, we can also use Equation \ref{B_alpha} and assume $f'=10$ to roughly estimate the toroidal magnetic field at the wind base \citep{Weiss-Bai-Fu-2021}. The resulting magnetic field strength range assumed in our simulations are approximately $B_{\rm mid} \simeq [0.40,12.56]$ Gauss ($m=10$, when the net vertical field is aligned with the disc rotation) or $B_{\rm mid} \simeq [0.04,1.26]$ Gauss ($m=1$, anti-aligned cases).

For a system with a host star mass of 1 $M_\odot$, the median accretion rate during the lifetime of the solar nebula is roughly estimated to be $10^{-8}$ $M_\odot \text{yr}^{-1}$ (i.e., during the Class II phase). The upper limit in our model ($2\times10^{-6}$ $M_\odot\rm yr^{-1}$) corresponds to typical values for Class 0 discs. For regions within 100 au of PPDs, a magnetic field strength range of 0.01 to 1 Gauss is considered reasonable. 
In areas very close to the center star or during the Class 0 phase, the magnetic field can exceed 10 Gauss. 
Therefore, the assumptions and parameter space in our model are consistent with realistic PPD environment.

From Figure \ref{fig:rtatorque_radial}, we confirm that the transition from inward to outward migration towards positive torque 
(in the case of low $q$, minimum negative torque) is consistent with expectation across different planet masses. 
More quantitatively, 
one can define $K  = \tau_{\rm dw}/\tau_{\rm lib} \sim (q/h_0^3) \alpha_{\rm dw}^{-1}$ similar to \citetalias{kimmig2020} and this transition occurs at $K\lesssim 10$. Our findings of the transition from inward to outward migration at $K = 10$ 
agree with the findings in \citetalias{kimmig2020}. 

The transition from outward to inward migration that we emphasize in this letter, 
where the maximum horseshoe drag relaxes to the linear corotation torque, 
is consistent with $K\lesssim 10 h_0$. 
This is when the radial advection timescale is even shorter than the U-turn timescale $ \tau_{\rm U-turn} \sim \tau_{\rm lib} h_0$. 
Note that the half-width of the "bumps" in the torque profiles shown in 
Figure \ref{fig:rtatorque_radial} are all roughly a factor of $\log (1/h_0)$, 
insensitive to the planet mass. 

\begin{figure}
\centering
\includegraphics[width=0.99\hsize]{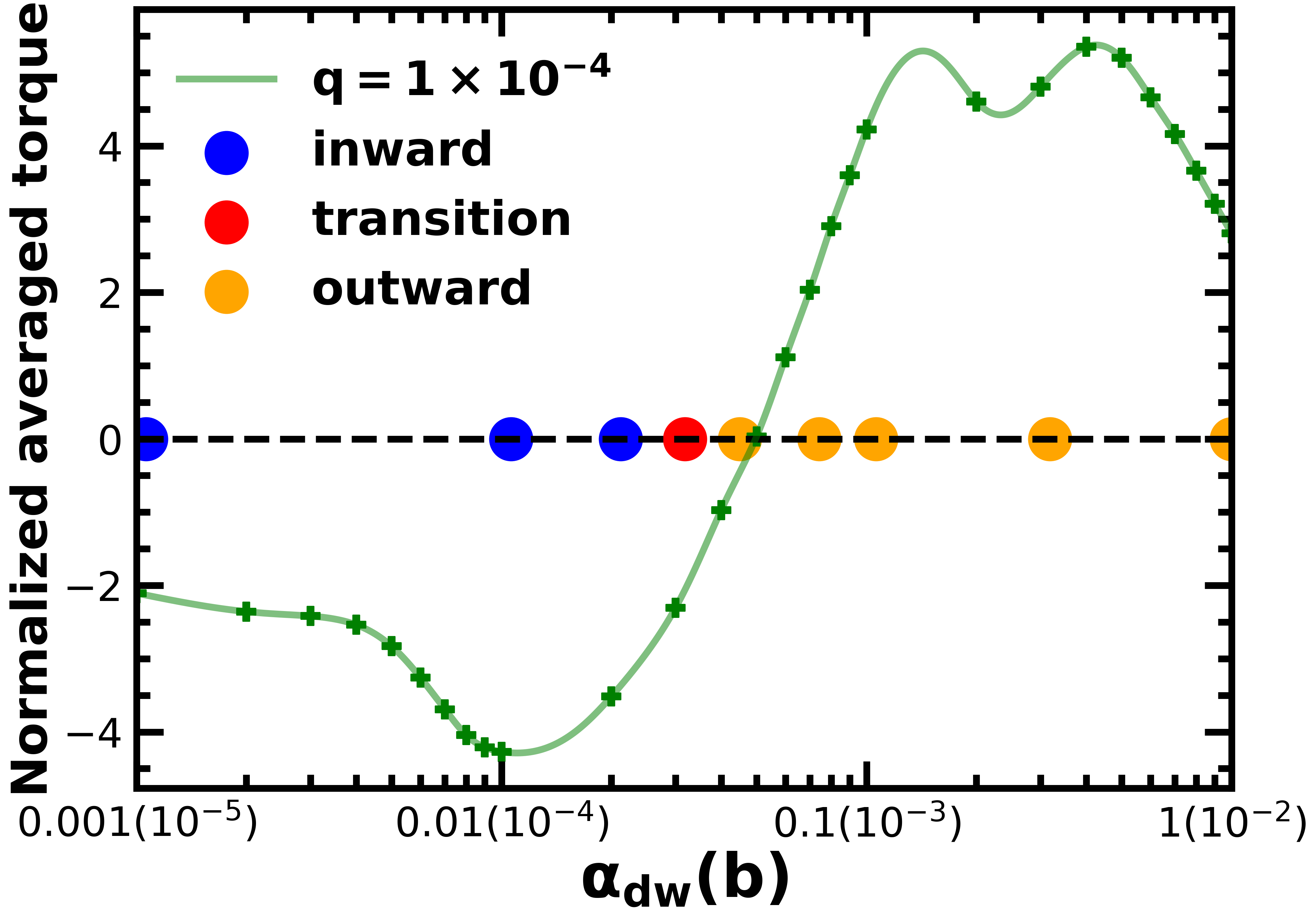}
\caption{Similar to the left panel of Figure \ref{fig:rtatorque_radial}, 
but with the addition of the value of $b$ on the x-axis corresponding to the definition of \citetalias{kimmig2020}, which is roughly a factor of 100 smaller than $\alpha_{\rm dw}$ for $h_0=0.05$. 
The green line represents cases with $q= 1\times10^{-4}$, a marginally gap-opening companion ($q\sim h_0^3$). 
The colored dots indicate the results 
from Figure 8 in \citetalias{kimmig2020} for Saturn case. 
Note that the planet mass in \citetalias{kimmig2020} is $q=3 \times 10^{-4}$, while we use $q=1 \times 10^{-4}$. Due to differences in the simulation domain and resolution, there may be other additional slight discrepancies in the results.}\label{fig:kimmig}
\end{figure}

\subsection{The Gap-opening Limit}

To connect with the parameter space for more massive planets, 
we perform another simulations with $q = 10^{-4}$, 
a mass ratio that is marginally super-thermal for $h_0 = 0.05$ ($q\sim h_0^3$). 
In this case, the torque dependence becomes more non-linear due to gap-opening effects.
While the upper limit for $\alpha_{\rm dw}$ that allows for outward migration still exists, it already exceeds $\alpha_{\rm dw}\sim 1$, 
which is too large to be realistic if $\alpha_{\rm dw}$ is a global constant unless $h_0$ is exceptionally large. 
For comparison, 
we plot \citetalias{kimmig2020} results for a Saturn-like planet with $q\sim 3\times 10^{-4}$ along a straight line (since only the direction but not the magnitude of the torque is given in their figure), 
relating their accretion parameter $b$ with our $\alpha_{\rm dw}$.


Nevertheless, when the planet is massive enough to carve out a gap in the disc, MHD simulations by \citet{AoyamaBai2023} indicate that the effective wind accretion parameter can be locally quite large within 
the gap due to magnetic flux concentration. 
They also typically observe that the horseshoe libration region 
is strongly compressed by radial advection velocities, 
similar to the lower panel of our Figure \ref{fig:streaming_line_alpha_dw} (see their Figure 12) when wind dominates disc accretion. 
This suggests that if the effect of locally large $\alpha_{\rm dw}$ shifting the corotation torque toward the linear regime is properly considered, 
the migration direction for gap-opening planets may still be typically inwards.

\section{Discussion}

In this paper, 
we investigated the pace and direction of planet migration in a PPD dominated by 
wind-driven accretion instead of turbulence-driven accretion. 
Building on the works of \citetalias{kimmig2020}, 
we run simulations across a larger parameter space, 
covering lower-mass planets and higher accretion rates. 
We confirm that due to the advection of materials across the horseshoe, 
when $\tau_{\rm dw}  \lesssim  \tau_{\rm lib}$, a non-linear horseshoe drag is produced 
which reduces the absolute value of the negative torque, 
and can 
sometimes lead to a reversal in the torque sign.
The magnitude of this reversal, 
normalized by the linear Lindblad torque, is observed to be more pronounced for higher-mass planets (see Figure \ref{fig:rtatorque_3cases}). In addition, 
we find that when $\tau_{\rm dw}$ is shorter than the U-turn timescale $\sim \tau_{\rm lib} h$, 
the horseshoe drag will relax to the linear corotation torque, 
which usually does not dominate over the Lindblad torque \citep{Muto-Inutsuka2009} so the planet is in a state of inward migration. 
Therefore, 
the outward migration may occupies only a limited parameter space where horseshoe drag is close to its maximum non-linear value. 
We emphasize that both transitions 
are analogous to the well-studied 
corotation torque desaturation phenomenon in classical type I migration for viscous discs \citep{PaardekooperPap2009,Paardekooper2011}, 
although net viscous diffusion across the gap 
occurs only in a statistically-averaged sense when a reasonable surface density gradient is present.

The implications of our finding 
are 
particularly interesting for moderately massive planets 
where $q$ is large enough for its maximum horseshoe drag to dominate over the Lindblad torque (e.g., the blue and yellow curves in Figure \ref{fig:rtatorque_radial}). 
Since the parameter $K = q \alpha_{\rm dw}^{-1} h^{-3}$ is sensitive to the aspect ratio $h$, which is typically an increasing function of radius, 
a planet will migrate outwards if formed outside a critical radius $r_1$ that satisfies $h^3(r_1)\sim (q \alpha_{\rm dw}^{-1}/10)$. 
Moreover, beyond some second radius $r_2$ where $h^4(r_2)\sim (q \alpha_{\rm dw}^{-1}/10)$, the second transition occurs, causing the planet's orbital evolution to stall close to $r_2$. If we adopt the minimum-mass solar nebula model \citep{ChiangYoudin2010, Piso2014, Chen2020a} where $h \approx 0.05 (r/10{\rm au})^{2/7}$, we have 

\begin{equation}
    r_1 \approx 7.7 {\rm au} \left(\dfrac{q}{10^{-5}} \dfrac{10^{-2}}{\alpha_{\rm dw}}\right)^{7/6}, r_2 \approx 113.1 {\rm au} \left(\dfrac{q}{10^{-5}} \dfrac{10^{-2}}{\alpha_{\rm dw}}\right)^{7/8}.
\end{equation}

In the absence of a migration trap, 
the planet could continue migrating towards the edge of the PPD due to the non-linear horseshoe drag. 
Therefore, 
the trap at $r_2$ plays a role in retaining intermediate-mass or massive planets at a few tens or hundred au, 
which could be responsible for the multi-ringed structures observed by ALMA \citep{Huang2018}.
Here a wind accretion parameter of $\alpha_{\rm dw} = 10^{-2}$ corresponds to 
an accretion rate of $\dot{M} \sim 6\times 10^{-8} M_\odot{\rm yr}^{-1}$ and field strength of $B\sim 0.2-2~{\rm Gauss}$ (see discussion in \S \ref{sec:Psurvey}), which
falls within the typical values for Class II discs \citep{Weiss-Bai-Fu-2021}, 
indicating that this conclusion is applicable to planet-forming discs. 

Our results can also be related to 
migration under effect of other physical processes that generally involve advection of materials across the horseshoe region. 
For example, the Hall effect can also lead to an inward radial velocity \citep{WuLin2024},   
the influence of which on planet migration is analyzed in \citet{McNally2017, McNally2018}. 
In fact, 
they also generally define a $\chi$ parameter as the ratio between the radial
"flushing" timescale and the libration timescale, similar to $K$ defined in \S \ref{sec:Psurvey} or in \citetalias{kimmig2020}, 
and show there is a growing influence of the unbalanced horseshoe drag $\chi$ 
approaches order unity from infinity 
(in other words, for increasing radial advection velocity). 
In their paper, they also suggest that in the limit of $\chi \ll 1$,
the corotation torque may eventually enter the linear regime, though they did not explicitly demonstrate this. 
Our simulation results provide direct evidence supporting this hypothesis. Nevertheless, 
turbulence from surface active layers may still exist in realistic magnetized discs \citep{Gammie1996, Lega-2022} to interact with MHD wind effects, an extra complexity that our analysis does not account for.

Finally, we note that the dust torque discussed in \citet{Llambay2018, Guilera2023,Hou-Yu-2024,Chrenko-2024} may also fit within the general framework described above, or at least partially.
Dust particles rotating on Keplerian azimuthal velocities in the disc experience friction against the sub-Keplerian gas, 
resulting in a drift velocity $v_{\rm dust}$ \citep{Weidenschilling1977}. 
This drift can lead to an overdensity of dust in the upper horseshoe region and a deficit in the lower horseshoe, when the drift timescale is comparable to the libration timescale. 
However, $v_{\rm dust}$
is influenced by various non-linear factors in the presence of a planet, 
including highly non-axisymmetric pressure gradients and changes in the Stokes number 
caused by density waves and gap-opening. 
Therefore, accurately capturing transitions in the dust torque profile with a single parameter like $K$ or $\chi$ can be challenging, warranting further investigation. 


\section*{Acknowledgements}
We thank the anonymous referee for a thorough review and highly constructive suggestions, which significantly enhanced the quality of the manuscript. We also thank Richard Alexander, Xue-Ning Bai, Cl\'ement Baruteau, Richard Nelson and Sijme-Jan Paardekooper for discussions and useful suggestions. This research used the DiRAC Data Intensive service at Leicester, operated by the University of Leicester IT Services, which forms part of the STFC DiRAC HPC Facility (\href{www.dirac.ac.uk}{www.dirac.ac.uk}).

\section*{Data availability}
The data obtained in our simulations can be made available on reasonable request to the corresponding author. 

\appendix

\bibliographystyle{mnras}
\bibliography{windy-disc}

\end{CJK*}
\bsp	
\label{lastpage}

\end{document}